# Thermal tomography of asteroid surface structure

Alan W. Harris and Line Drube

German Aerospace Center (DLR) Institute of Planetary Research, Rutherfordstrasse 2, 12489 Berlin, Germany

alan.harris@dlr.de

ABSTRACT

Knowledge of the surface thermal inertia of an asteroid can provide insight into surface structure: porous material has a lower thermal inertia than rock. We develop a means to estimate thermal inertia values of asteroids and use it to show that thermal inertia appears to increase with spin period in the case of main-belt asteroids (MBAs). Similar behavior is found on the basis of thermophysical modeling for near-Earth objects (NEOs). We interpret our results in terms of rapidly increasing material density and thermal conductivity with depth, and provide evidence that thermal inertia increases by factors of 10 (MBAs) to 20 (NEOs) within a depth of just 10 cm. Our results are consistent with a very general picture of rapidly changing material properties in the topmost regolith layers of asteroids and have important implications for calculations of the Yarkovsky effect, including its perturbation of the orbits of potentially hazardous objects and those of asteroid family members after the break-up event. Evidence of a rapid increase of thermal inertia with depth is also an important result for studies of the ejecta-enhanced momentum transfer of impacting vehicles ("kinetic impactors") in planetary defense.

1. INTRODUCTION

The thermal inertia of an asteroid's surface provides a guide to its porosity and cohesion. Low values of thermal inertia are consistent with a dusty, porous regolith, such as that of the Moon, while high values are indicative of low-porosity rocky material with relatively high thermal conductivity. The Yarkovsky effect causes a secular variation in asteroid semimajor axis and is therefore an important effect in calculations of impact probabilities of potentially hazardous asteroids (see, e.g., Bottke et al. 2006; Vokrouhlický et al. 2015). The magnitude of the Yarkovsky effect, which arises as a result of anisotropic thermal emission, depends on the surface temperature distribution on an asteroid, which in turn depends on the thermal conductivity and thermal inertia of the surface material. Furthermore, some knowledge of the near-surface internal structure of a target asteroid is crucial for accurate predictions of the outcome of a deflection attempt with a kinetic impactor or explosion. In the case of an impact that produces a large mass of ejecta in the backward direction, there is significant enhancement of the forward momentum of the asteroid. For example, a bare-rock surface would be expected to lead to a greater momentum boost than a porous dusty surface (Jutzi et al. 2015).

Calculating accurate values of thermal inertia for asteroids is a difficult process requiring a shape model, thermal-infrared observations of the object obtained over broad ranges of rotation period and aspect angle, and detailed thermophysical modeling. Consequently, reliable thermal inertia values are currently available for relatively few asteroids (Delbo' et al. 2015). Thermophysical models include phenomena such as thermal inertia, surface roughness, and rotational state explicitly as model parameters and normally provide more accurate results than simple models based on spherical geometry. However, the extra complexity of thermophysical models is rarely warranted unless an accurate shape model of the object is available. On the basis of relatively simple asteroid thermal modeling we develop an empirical relationship enabling the thermal inertia of an asteroid to be roughly estimated given adequate measurements of its thermal-infrared continuum and knowledge of its spin vector. Use of the relationship with data from the NASA Wide-Field Infrared Survey Explorer space telescope (*WISE*, Wright et al. 2010) suggests that the thermal inertia of MBAs increases with decreasing spin rate. Furthermore, we demonstrate that published thermal inertia values of NEOs from thermophysical modeling show a similar dependence on spin rate. An explanation for spin-rate-dependent thermal inertia in terms of the thermal skin depth and depth-dependent thermal properties is provided. Our discovery of spin-rate-dependent thermal inertia throughout the asteroid population has important implications for the physical properties of the near-surface material layers.

## 2. ESTIMATING ASTEROID THERMAL INERTIA

### *2.1 Thermal Modeling*

The Near-Earth Asteroid Thermal Model (NEATM, Harris 1998) is a simple asteroid thermal model based on spherical geometry, which offers a means of deriving diameters and albedos of asteroids and trans-Neptunian objects (TNOs) (in principle all atmosphereless bodies) from thermal-infrared data for objects with unknown physical characteristics. The model incorporates a fitting parameter, $\eta$ that allows the model surface temperature distribution to be adjusted to take account of the effects of thermal inertia, spin vector, and surface roughness, thereby giving a better fit of the model fluxes to the measurements. The adjusted subsolar equilibrium temperature, $T_{SS}$, is proportional to $\eta^{-1/4}$; the temperature distribution over the illuminated surface of the model is given by $T_{SS} \cos^{1/4}\varphi$, where $\varphi$ is the angular distance of the surface element from the subsolar point. A rough surface gives rise to a higher measured subsolar temperature than expected for a smooth surface, due to enhanced emission from surface elements that happen to be facing the Sun (the "beaming" effect). In general, a rotating spherical asteroid with non-zero thermal inertia will have a cooler subsolar region and a longitudinal temperature distribution on the "evening" side that does not fall to zero at the terminator. The effects of a rough surface act so as to reduce $\eta$, while those of thermal inertia and rotation cause $\eta$ to increase. The best-fit $\eta$ value is a measure of the departure of an asteroid's temperature distribution from that of an object with a smooth surface and zero thermal inertia, or zero spin, in thermal equilibrium with insolation (in which case $\eta$ = 1). For further details of the NEATM see, e.g., Harris (1998), Delbo' & Harris (2002), Harris & Lagerros (2002). A number of investigators (e.g., Delbo' et al. 2003, 2007; Harris 2006; Wolters et al. 2005; Emery & Lim 2011; Lellouch et al. 2013; Harris & Drube, 2014) have used best-fit $\eta$ values from the NEATM as a guide to thermal inertia, defined as $\Gamma = (\kappa\rho c)^{0.5}$, where $\kappa$ is the thermal conductivity, $\rho$ the density, and $c$ the specific heat of the material.

If the spin vector is known, best-fit values of $\eta$ can be used as a rough proxy for the thermal parameter (Spencer et al. 1989),

$$\Theta = \Gamma\omega^{0.5}(\varepsilon\sigma T_{SS}^3)^{-1}, \tag{1}$$

where $\omega$ is the spin frequency, $\varepsilon$ the bolometric emissivity (normally assumed to be 0.9), $\sigma$ the Stefan Boltzmann constant, and $T_{SS}$ is given by $(S_1(1-A)/\eta\varepsilon\sigma R^2)^{1/4}$, in which $S_1$ is the solar constant, $A$ the bolometric Bond albedo, and $R$ the heliocentric distance in AU. Values of $\eta$ derived from thermal-infrared observations range from about 0.5 for an object with a rough surface and very low thermal inertia, in which case $\Theta \sim 0$, to around 3.0 for a rapidly spinning object with very high thermal inertia, in which case $\Theta \gg 1$. The thermal parameter and $\eta$ both increase with increasing thermal inertia and increasing spin rate. By setting $T_{SS}$ equal to the subsolar equilibrium (i.e. $\eta$ = 1) temperature, $\Theta$ is defined as a property of the body itself (Spencer et al. 1989; see Fig. 1). In contrast, $\eta$ depends on the spin-axis orientation with respect to the solar direction (the solar aspect angle, $\theta$), which can be calculated from the spin vector of the observed asteroid and its heliocentric coordinates at the time of the observation.

Fitted values of $\eta$ are now available for thousands of asteroids, thanks to the productivity of projects such as WISE/NEOWISE. However, spin vectors are available for a growing but still severely limited number of asteroids. Much effort was invested in gathering the dates, times and geometric data of *WISE* observations of those asteroids with known spin vectors (Warner et al. 2009). A survey of the sizes and albedos of more than 100,000 asteroids has been carried out by *WISE*, which was launched to Earth orbit in December 2009 carrying a 40-cm-diameter telescope and infrared detectors. The *WISE*/NEO*WISE* program (Masiero et al. 2011; Mainzer et al. 2011b) analyzed images collected during the cryogenic phase of the mission in up to four infrared bands, centered on 3.4, 4.6, 12, and 22 μm, and used a NEATM fitting routine to derive asteroid diameters, albedos, and corresponding best-fit $\eta$ values. The dates and times of the *WISE* observations of a particular target were accessed via the NASA/IPAC Infrared Science Archive General Catalog Query Engine using a script especially written for the purpose. The dates and times enabled ephemerides of the target to be generated for the times of the *WISE* observations via another purpose-written script by the NASA/JPL Solar System Dynamics Horizons Web-Interface. Knowledge of the ephemerides, in particular heliocentric ecliptic coordinates, allowed the solar aspect angle, $\theta$, to be calculated.

## 2.2. The Dependence of $\eta$ on Solar Aspect Angle, $\theta$

If an asteroid's spin axis is oriented close to the solar direction, the surface temperature distribution and $\eta$ will be largely independent of thermal inertia and spin rate. A simple illustration of the dependence of the NEATM fitting parameter, $\eta$, on $\theta$ is given in Fig. 1. The diagram shows how, for a model asteroid, the temperature distribution on the surface of the asteroid varies with decreasing solar aspect angle, leading to a decrease in $\eta$, which reaches a minimum at $\theta = 0°$. Therefore values of $\eta$ derived from observational data should depend on the spin-axis orientation with respect to the solar direction. Using fitted $\eta$ values from Masiero et al. (2011) for MBAs with known spin vectors, we have confirmed that there is a general decrease in $\eta$ with decreasing $\sin\theta$ (Fig. 2), and therefore that $\eta$ carries useful information on the surface temperature distribution.

## 2.3. The Dependence of $\eta$ on the Thermal Parameter, $\Theta$

A convenient table of published thermal inertia values for some 50 asteroids, including NEOs, main-belt asteroids, and TNOs, derived from detailed thermophysical modeling is provided by Delbo' et al. (2015). For most of the NEOs in the list of Delbo' et al. (2015) we have obtained $\eta$ values, either directly from the literature or by calculating them by applying the NEATM to thermal-infrared flux measurements taken from the literature or from the *WISE* data archive. Uncertainty estimates for derived values of $\eta$ differ considerably, depending on the source. Some authors give formal statistical $1\sigma$ error bars while others quote larger, more conservative, error bars, which reflect experience of the repeatability of $\eta$ values derived from different sets of measurements. We have taken a conservative approach and assigned error bars of $\pm$ 20% to $\eta$, except where those from the original source are larger. In each case the thermal parameter was calculated, and the solar aspect angle, $\theta$, was derived from the ecliptic coordinates of the NEO's spin vector and its heliocentric ecliptic coordinates at the time of the observations. In order to remove the dependency of $\eta$ on solar phase angle, $\alpha$, we have used the linear relation derived by Mainzer et al. (2011b; see their fig. 7) to normalize the $\eta$ values to $\alpha = 50°$, which is near the center of the range at which *WISE* observations of NEOs are made. A plot of normalized $\eta$ versus the thermal parameter multiplied by $\sin\theta$ (Fig. 3) shows a very significant positive correlation, which has not been previously demonstrated. The data used in Fig. 3 are listed in Table 1.

We have chosen NEOs for the purposes of Fig. 3 due to the fact that compared to MBAs their $\eta$ values cover a much larger range, and observational circumstances, such as aspect angle range, facilitate more reliable thermophysical modeling. Furthermore, infrared observations of large MBAs may suffer from saturation issues (Masiero et al. 2011) and consequently provide less reliable $\eta$ values. The theoretical relationship between $\eta$ and the thermal parameter, as given on the basis of a thermophysical model, is not linear (cf. Spencer et al. 1989, fig. 5; Lellouch et al. 2013, fig. 5). In Fig. 3 we have superimposed a curve to illustrate the form of the theoretical dependence of $\eta$ on $\Theta$, which rises asymptotically to a maximum as $\Theta$ tends to infinity. While the theoretical relationship between $\eta$ and $\Theta$ is more complex, it is evident from Fig. 3 that the weighted linear fit to the data (continuous line) serves as a very good approximation, at least in the range $0.75 < \Theta \sin\theta < 3.5$ covered by the data.

*2.4. A NEATM-based Thermal Inertia Estimator*

The best weighted linear fit to the data of Fig. 3 is given by $\eta_{norm} = 0.74 + 0.38 \times \Theta \sin\theta$. Substituting for $\Theta$ (Equation 1), and $T_{SS}$ (see above), we derive an expression for estimating thermal inertia, given a measurement of $\eta$:

$$\Gamma = (\eta_{norm} - 0.74)((1-A)^{3/4} \; \varepsilon^{1/4} \; \sigma^{1/4} \; S_I^{3/4})/(0.38 \sin\theta \; \omega^{1/2} \; R^{3/2}) \quad \text{J m}^{-2}\text{s}^{-0.5}\text{K}^{-1}, \qquad (2)$$

where $\eta_{norm} = \eta + (50° - \alpha°) \times 0.00963$ (Mainzer et al. 2011b). There is little evidence that $\eta$ depends on $\alpha$ for $\alpha < 20°$ (Mainzer et al. 2011b), therefore for the purposes of normalization we take $\alpha = 20°$ for phase angles below 20°. Since the Sun's spectral energy distribution peaks in the visible region and the dependence of asteroid albedos on wavelength is normally small, it is usual to assume that $A = A_V = q \; p_V$, where $q$ is the phase integral. This allows the physically significant parameter $A$ to be linked directly to the observationally derived visible geometric albedo, $p_V$. In the standard $H$, $G$ magnitude system (Bowell et al. 1989), in which $H$ is the absolute magnitude and $G$ is the slope parameter, $q = 0.290 + 0.684 \; G$.

To test the reliability of our NEATM-based thermal-inertia estimator we have used it to estimate thermal inertia for objects with known thermal inertia values derived via thermophysical modeling (Delbo' et al. 2015), for which reliable values of the required parameters were accessible from the literature or could be derived from on-line databases. We have excluded objects for which $\Theta \times \sin\theta$ is outside the range 0.75 – 3.5 (see above). A comparison of estimated thermal inertia values with those derived by means of thermophysical modeling is shown in Fig. 4. The value of the thermal-inertia estimator is immediately evident, especially given that Fig. 4 includes not only the near-Earth asteroids in Fig. 3 but also main-belt asteroids, Centaurs, and trans-Neptunian objects. In nearly all cases the $\eta$-based estimates agree within the error bars (see below) with the values derived from thermophysical modeling. Note that the plot covers nearly 4 orders of magnitude of thermal inertia. The RMS fractional difference between the $\eta$-based estimates and the values derived from thermophysical modeling is 40.0%. The data used in Fig. 4 are listed in Table 1.

We have also investigated the expected accuracy of Equation 2 by considering three sources of uncertainty:

1. The slope of the weighted best-fit relationship between $\eta$ and solar phase angle, $\alpha$, given by Mainzer et al. (2011b) is 0.00963 ± 0.00015. We find that the quoted uncertainty has a negligible effect on the slope of the linear fit of Fig. 3, modifying the numerical constants in Equation 2 only in the third decimal place. We have investigated how the uncertainty in the Mainzer et al. (2011b) relationship propagates through to the thermal inertia values based on Equation 2 given in Table 1: again the results reveal an insignificant RMS fractional error of 0.82%.
2. The slope of the weighted linear fit in Fig. 3 has 1 $\sigma$ uncertainties shown in the plot as dashed lines; propagation through Equation 2 to the thermal inertia values in Table 1 results in a RMS fractional contribution to the error budget of 17%.
3. We assume a conservative overall uncertainty of 20% in measured $\eta$ values, as explained above, which contributes 49.7% to the error budget. This contribution is by far the most significant.

Adding the above three uncertainty contributions in quadrature gives an overall uncertainty of 52.5% in thermal inertia values estimated via Equation 2, which is larger than the results from the comparison of thermal inertia values estimated via Equation 2 with those derived via detailed thermophysical modelling discussed above. The fact that the theoretical error budget is somewhat larger may be due to our assumption of a 20% uncertainty in $\eta$ being over-conservative.

As mentioned in Section 2.1 best-fit $\eta$ values derived via use of the NEATM depend on surface roughness in addition to thermal inertia. A linear fit derived from a set of objects having different surface roughness characteristics to the objects plotted in Fig. 3 would presumably have a modified intercept and slope (e.g., see Lellouch et al. 2013, fig. 5), with rougher/smoother surfaces giving rise to a slightly steeper/shallower slope. Implicit in the use of Equation 2 is the assumption that the surface roughness of the asteroid in question is compatible with the slope of the linear fit in Fig. 3. In any case, the error analysis above indicates that the effects of different degrees of roughness are adequately accounted for in the conservative 20% uncertainty assumed for measured values of $\eta$, at least for the set of objects in Table 1.

Another source of uncertainty in $\eta$ is the effect of shape. For highly irregular objects the temperature and therefore $\eta$ value observed depend on rotational phase. Brown (1985) showed that in cases of marked departure from sphericity, use of simple models can give rise to significant errors due to differences in temperature distributions between ellipsoids and spheres. The overall effect is an increase in thermal lightcurve amplitude above that expected from the variation in projected area due to rotation. In practice, however, the error can be minimized by taking measurements at several points on the lightcurve, which is often done in thermal-infrared observations of asteroids. For example, the *WISE* cryogenic survey made an average of 10 detections of a typical asteroid or comet spaced over ~ 36 h (Mainzer et al. 2011a). Again, the error analysis above indicates that the combined effects of factors influencing $\eta$ are adequately accounted for in the conservative 20% uncertainty assumed for measured values of $\eta$.

Of course, individual objects may have physical characteristics differing considerably from those of the population in general, or the combination of observing geometry, thermal inertia, and rotation rate may conspire to give a significant "morning/evening" effect, which would increase/decrease $\eta$ (an extreme case appears to be the two entries for 1998 WT$_{24}$ in Table 1; see Harris et al. 2007). In such cases results obtained from Equation 2 may be less accurate. However, occasional anomalous results will not affect the conclusions of this study, which is based on observations of hundreds of objects. Further data on asteroid thermal properties and detailed thermophysical modeling will enable the parameters in Equation 2 to be optimized and the scope of its usefulness to be better defined.

**Table 1**
Data used in Figs. 3 and 4

| Name | G | R (AU) | Period (h) | $p_V$ | $\alpha°$ | $\sin\theta$ | $\eta$ | $\eta_{err}$ | $\eta_{norm}$ | $\eta_{norm\,err}$ | $\Theta\sin\theta$ | $\Gamma$ | $\Gamma_{err}$ | $\Gamma_{TP}$ | $\Gamma_{TP\,err}$ | Notes |
|---|---|---|---|---|---|---|---|---|---|---|---|---|---|---|---|---|
| **NEOs** | | | | | | | | | | | | | | | | |
| 433_Eros | 0.15 | 1.62 | 5.27 | 0.21 | 31.0 | 0.82 | 1.07 | 0.20 | 1.26 | 0.24 | 1.46 | 138 | 68 | 150 | 50 | 1 |
| 433_Eros | 0.15 | 1.13 | 5.27 | 0.20 | <20 | 0.99 | 1.05 | 0.11 | 1.34 | 0.14 | 1.02 | 229 | 103 | 150 | 50 | 2 |
| 1580_Betulia | 0.15 | 1.14 | 6.13 | 0.11 | 53.0 | 0.97 | 1.09 | 0.22 | 1.06 | 0.21 | 1.10 | 136 | 92 | 180 | 50 | 3 |
| 1862_Apollo | 0.15 | 1.1 | 3.06 | 0.26 | 35.3 | 0.97 | 1.15 | 0.23 | 1.29 | 0.26 | 1.20 | 168 | 79 | 140 | 100 | 2 |
| 25143_Itokawa | 0.21 | 0.98 | 12.13 | 0.19 | 109.0 | 1.00 | 2.85 | 0.57 | 2.28 | 0.46 | 2.58 | 1094 | 324 | 700 | 200 | 4 |
| 33342_1998_WT24 | 0.15 | 1.01 | 3.70 | 0.56 | 60.4 | 0.88 | 1.86 | 0.38 | 1.76 | 0.36 | 1.40 | 381 | 135 | 200 | 100 | 5 |
| 33342_1998_WT24 | 0.15 | 0.99 | 3.70 | 0.56 | 79.3 | 0.96 | 1.25 | 0.25 | 0.97 | 0.19 | 1.45 | 81 | 70 | 200 | 100 | 5 |
| 99942_Apophis | 0.24 | 1.04 | 30.56 | 0.30 | 60.4 | 0.99 | 1.82 | 0.37 | 1.72 | 0.35 | 1.56 | 982 | 351 | 600 | 300 | 6 |
| 101955_Bennu | 0.15 | 1.13 | 4.30 | 0.05 | 63.4 | 1.00 | 1.55 | 0.03 | 1.42 | 0.03 | 2.23 | 246 | 103 | 310 | 70 | 7 |
| 162173_Ryugu | -0.115 | 1.29 | 7.63 | 0.07 | 22.3 | 0.99 | 1.83 | 0.37 | 2.10 | 0.42 | 2.62 | 541 | 169 | 400 | 200 | 8 |
| 162173_Ryugu | -0.115 | 1.2 | 7.63 | 0.07 | 52.6 | 0.97 | 1.63 | 0.15 | 1.61 | 0.15 | 2.30 | 392 | 146 | 400 | 200 | 9 |
| 175706_1996_FG3 | -0.041 | 1.23 | 3.59 | 0.04 | 54.9 | 1.00 | 1.27 | 0.25 | 1.22 | 0.24 | 1.07 | 142 | 72 | 120 | 50 | 10 |
| 175706_1996_FG3 | -0.041 | 1.38 | 3.59 | 0.04 | <20 | 1.00 | 1.15 | 0.23 | 1.44 | 0.29 | 1.27 | 172 | 71 | 120 | 50 | 11 |
| 341843_2008_EV5 | 0.15 | 1.03 | 3.72 | 0.11 | 73.0 | 1.00 | 2.04 | 0.44 | 1.82 | 0.39 | 3.14 | 404 | 147 | 450 | 60 | 10 |
| 308635_2005_YU55 | -0.13 | 0.99 | 19.31 | 0.06 | 34.0 | 1.00 | 1.08 | 0.22 | 1.23 | 0.25 | 1.68 | 442 | 226 | 575 | 225 | 12 |
| 308635_2005_YU55 | -0.13 | 0.99 | 19.31 | 0.06 | 34.0 | 0.87 | 1.08 | 0.21 | 1.23 | 0.24 | 1.47 | 505 | 253 | 575 | 225 | 12 |
| **MBAs + others** | | | | | | | | | | | | | | | | |
| 16_Psyche_I | 0.15 | 3.16 | 4.20 | 0.12 | <20 | 0.49 | 0.86 | 0.07 | 1.15 | 0.10 | 2.16 | 62 | 35 | 125 | 40 | 13 |
| 22_Kalliope_I | 0.21 | 3.07 | 4.15 | 0.15 | <20 | 0.19 | 0.75 | 0.03 | 1.04 | 0.04 | 0.82 | 118 | 83 | 125 | 125 | 13 |
| 22_Kalliope_W | 0.21 | 2.93 | 4.15 | 0.17 | 20.3 | 0.82 | 1.08 | 0.05 | 1.37 | 0.06 | 3.29 | 62 | 27 | 125 | 125 | 14 |
| 32_Pomona_I | 0.15 | 2.81 | 9.45 | 0.23 | 21.2 | 0.98 | 0.98 | 0.12 | 1.26 | 0.15 | 1.40 | 67 | 33 | 70 | 50 | 13 |
| 32_Pomona_W | 0.15 | 2.39 | 9.45 | 0.25 | 24.8 | 1.00 | 1.05 | 0.14 | 1.29 | 0.17 | 1.12 | 90 | 42 | 70 | 50 | 14 |
| 44_Nysa_I | 0.15 | 2.44 | 6.42 | 0.47 | 24.0 | 1.00 | 1.04 | 0.06 | 1.29 | 0.07 | 2.60 | 67 | 31 | 120 | 40 | 13 |
| 87_Sylvia_I | 0.15 | 3.51 | 5.18 | 0.04 | <20 | 0.92 | 0.94 | 0.09 | 1.23 | 0.12 | 2.33 | 38 | 19 | 70 | 60 | 13 |
| 87_Sylvia_W | 0.15 | 3.31 | 5.18 | 0.05 | <20 | 0.90 | 0.86 | 0.01 | 1.15 | 0.01 | 2.10 | 35 | 20 | 70 | 60 | 14 |
| 107_Camilla_I | 0.08 | 3.71 | 4.84 | 0.04 | <20 | 0.90 | 0.97 | 0.09 | 1.26 | 0.11 | 0.91 | 37 | 18 | 25 | 10 | 13 |
| 107_Camilla_W | 0.08 | 3.74 | 4.84 | 0.06 | <20 | 0.98 | 0.99 | 0.10 | 1.28 | 0.13 | 1.01 | 35 | 17 | 25 | 10 | 14 |
| 110_Lydia_I | 0.2 | 2.88 | 10.93 | 0.16 | 20.7 | 0.96 | 0.98 | 0.14 | 1.26 | 0.18 | 2.51 | 73 | 36 | 135 | 65 | 13 |
| 110_Lydia_W | 0.2 | 2.9 | 10.93 | 0.17 | 20.3 | 0.94 | 0.97 | 0.08 | 1.26 | 0.10 | 2.48 | 73 | 36 | 135 | 65 | 14 |
| 115_Thyra_I | 0.12 | 2.47 | 7.24 | 0.25 | 24.0 | 0.87 | 0.94 | 0.08 | 1.20 | 0.11 | 1.03 | 71 | 37 | 62 | 38 | 13 |
| 121_Hermione_W | 0.15 | 3.29 | 5.55 | 0.08 | <20 | 1.00 | 1.06 | 0.13 | 1.35 | 0.17 | 0.96 | 49 | 22 | 30 | 25 | 14 |
| 130_Elektra_I | 0.15 | 3.53 | 5.22 | 0.07 | <20 | 1.00 | 0.94 | 0.05 | 1.22 | 0.07 | 1.09 | 35 | 18 | 30 | 30 | 13 |
| 130_Elektra_W | 0.15 | 3.11 | 5.22 | 0.09 | <20 | 0.98 | 0.94 | 0.04 | 1.23 | 0.05 | 0.90 | 42 | 21 | 30 | 30 | 14 |
| 277_Elvira_I | 0.15 | 2.63 | 29.69 | 0.20 | 22.5 | 1.00 | 1.10 | 0.19 | 1.37 | 0.23 | 2.57 | 159 | 70 | 250 | 150 | 13 |
| 277_Elvira_W | 0.15 | 3.14 | 29.69 | 0.20 | <20 | 1.00 | 0.92 | 0.01 | 1.21 | 0.01 | 3.35 | 91 | 47 | 250 | 150 | 14 |
| 283_Emma_I | 0.15 | 2.88 | 6.90 | 0.03 | <20 | 0.53 | 0.73 | 0.03 | 1.02 | 0.04 | 1.28 | 59 | 44 | 105 | 100 | 13 |
| 283_Emma_W | 0.15 | 2.77 | 6.90 | 0.03 | 20.7 | 0.62 | 0.84 | 0.01 | 1.12 | 0.01 | 1.43 | 73 | 43 | 105 | 100 | 14 |
| 306_Unitas_I | 0.15 | 2.17 | 8.74 | 0.19 | 27.9 | 0.91 | 1.01 | 0.14 | 1.22 | 0.17 | 2.33 | 97 | 49 | 180 | 80 | 13 |
| 306_Unitas_W | 0.15 | 2.71 | 8.74 | 0.20 | 21.5 | 0.85 | 0.85 | 0.02 | 1.12 | 0.03 | 3.03 | 59 | 35 | 180 | 80 | 14 |
| 382_Dodona_I | 0.15 | 2.56 | 4.11 | 0.13 | 23.3 | 0.86 | 1.09 | 0.17 | 1.34 | 0.21 | 1.79 | 70 | 31 | 80 | 65 | 13 |
| 382_Dodona_W | 0.15 | 2.76 | 4.11 | 0.14 | 21.3 | 1.00 | 1.58 | 0.02 | 1.86 | 0.02 | 2.34 | 100 | 33 | 80 | 65 | 14 |
| 694_Ekard_I | 0.15 | 1.84 | 5.93 | 0.04 | 33.4 | 0.85 | 0.95 | 0.11 | 1.11 | 0.13 | 1.30 | 89 | 53 | 120 | 20 | 13 |
| 720_Bohlinia_I | 0.15 | 2.89 | 8.92 | 0.14 | 20.3 | 0.86 | 1.11 | 0.17 | 1.40 | 0.21 | 2.46 | 94 | 40 | 135 | 65 | 13 |
| 956_Elisa_W11 | 0.15 | 2.76 | 16.49 | 0.15 | 21.1 | 0.91 | 0.99 | 0.04 | 1.27 | 0.05 | 1.19 | 104 | 50 | 90 | 60 | 15 |
| 1173_Anchises_W11 | 0.03 | 4.93 | 11.61 | 0.03 | <20 | 0.91 | 1.35 | 0.14 | 1.64 | 0.17 | 1.81 | 65 | 24 | 50 | 20 | 15 |
| 2060_Chiron | 0.15 | 14.87 | 5.92 | 0.18 | <20 | 0.91 | 0.91 | 0.10 | 1.20 | 0.13 | 1.39 | 4.3 | 2.3 | 5 | 5 | 16 |
| 10199_Chariklo | 0.15 | 13.51 | 7.00 | 0.04 | <20 | 0.91 | 1.12 | 0.08 | 1.41 | 0.10 | 3.40 | 8.2 | 3.5 | 16 | 14 | 16 |
| 90482_Orcus | 0.15 | 47.8 | 10.47 | 0.24 | <20 | 0.91 | 0.98 | 0.05 | 1.27 | 0.07 | 1.23 | 1.1 | 0.5 | 1 | 1 | 16 |
| 136108_Haumea | 0.15 | 51.1 | 3.92 | 0.80 | <20 | 0.91 | 0.95 | 0.33 | 1.24 | 0.43 | 0.82 | 0.5 | 0.4 | 0.3 | 0.2 | 16 |
| 208996_2003AZ84 | 0.15 | 45.4 | 6.78 | 0.11 | <20 | 0.91 | 1.05 | 0.20 | 1.34 | 0.26 | 1.63 | 1.2 | 0.5 | 1.2 | 0.6 | 16 |

**Notes to Table 1.** The column headed "$\Gamma$" contains values of thermal inertia (J m$^{-2}$s$^{-0.5}$K$^{-1}$) estimated from $\eta$ normalized to $\alpha = 50°$, as explained in the text. The column headed "$\Gamma_{TP}$" contains values of thermal inertia derived by means of detailed thermophysical modeling (Delbo' et al. 2015). The data set excludes objects with $\Theta$ sin$\theta$ outside the range 0.75 – 3.5 (see text), and those for which a reliable value of $\eta$ was not available or could not be calculated from the available data. The errors on the $\Gamma$ values result from the assumption of uncertainties in $\eta$ of at least ± 20% (see text). Spin vectors are from the sources cited or the Asteroid Lightcurve Database (Warner et al. 2009).

Data sources for the NEOs: 1. Harris & Davies (1999); 2. Harris (1998); 3. Harris et al. (2005); 4. Müller et al. (2005), $\eta$ calculation this work; 5. Harris et al. (2007), with spin vector from Busch et al. (2008), the observations were of the "evening" and "morning" sides of the object; 6. Müller et al. (2014), $\eta$ calculation this work (NEATM fit less secure, based on flux measurements on Rayleigh-Jeans side of thermal continuum only); 7. Emery et al. (2014); 8. Hasegawa et al. (2008), $\eta$ calculation this work; 9. Campins et al. (2009); 10. This work using data from the *WISE* cryogenic archive; 11. Wolters et al. (2011); 12. Müller et al. (2013), $\eta$ calculation this work, the two entries are for two possible pole solutions: for the purposes of Fig. 3 the mean of the corresponding two values of $\Theta$ sin$\theta$ has been taken.

Data sources for the MBAs and Jupiter Trojans: 13. For objects with names appended by "I" the source of the data is the IRAS SIMPS Catalog (Tedesco et al. 2002), $\eta$ values were derived in this work; 14. For objects with names appended by "W" the source of the data is the *WISE* catalog of Masiero et al. (2014); 15. For objects with names appended by "W11" the source of the data is the *WISE* catalog of Masiero et al. (2011).

Data source for the remaining objects: 16. Lellouch et al. (2013).

For objects with unknown pole directions (the last 7 objects in the table) $\theta = 65°$ (sin$\theta = 0.91$) was assumed, based on the mean value of sin$\theta$ for the objects with known $\theta$; for lower sin$\theta$ the resulting thermal inertia would be higher.

# 3. ASTEROID SPIN RATE AND THERMAL INERTIA

We have used the NEATM-based thermal-inertia estimator (Equation 2) to investigate the dependence of thermal inertia on asteroid rotation rate. In the case of MBAs we find an unexpected trend of increasing thermal inertia with decreasing spin rate (Fig. 5). Available thermal inertia values from thermophysical modeling (Delbo' et al. 2015) are overplotted in Fig. 5 (red points) on the estimated data from this work. While the thermophysically-modeled thermal inertia values taken on their own are insufficient in number and range of spin rate for any conclusion to be drawn on spin-rate-dependent thermal inertia, they do appear to be consistent with the trend apparent in the thermal inertia values estimated using Equation 2.

In the interests of an independent check on the behaviour of our thermal-inertia estimator, we investigated the dependence of $\eta$ on rotation period. The expected relation between $\eta$ and rotation period was calculated using a smooth-surface thermophysical model based on spherical geometry for a constant thermal inertia of 75 J m$^{-2}$s$^{-0.5}$K$^{-1}$ (cf. Fig. 2). The smooth-sphere thermophysical model, which is based on the work of Spencer et al. (1989), includes the effects of thermal inertia implicitly by determining the temperature of each surface element numerically via solution of the one-dimensional heat diffusion equation. The total observable thermal emission is calculated by summing the contributions from each surface element visible to the observer. The model infrared fluxes thus generated were fit by the NEATM to provide model $\eta$ values. As is evident in Fig. 6 the resulting model $\eta$ curve, which is normalized to the median measured $\eta$ value at rotation period = 4 h, does not fit the data for rotation period > 10 h and constant thermal inertia. As spin period increases the data points remain relatively high, while the model curve decreases, as expected for increasing period (Equation 1). The widening gap between the $\eta$ values and the model curve is consistent with increasing thermal inertia, as indicated by Equation 2 and evident in Fig. 5.

A similar trend to that in Fig. 5 has been noted by Delbo' et al. (2011) in the case of near-Earth asteroids, albeit of increasing $\eta$ values with increasing spin period, who suggested that it is possibly related to the YORP effect (e.g. Vokrouhlický et al. 2015), i.e. modification of the spin rate of an irregularly shaped body via the reflection and thermal re-emission of solar radiation. YORP acts on small asteroids more effectively than on large asteroids, therefore it is questionable whether YORP could also explain the slow rotators with high thermal inertia in Fig. 5. A mechanism has been discussed by Pravec et al. (2002) by which the asymmetric distribution of escaping ejecta from asteroids with diameters of ~100 km may lead to a slowing of the spin rate with many impacts. However, it is not clear how a relatively large impact rate could lead to higher thermal inertia.

We investigated the dependence of the thermal inertia values of NEOs derived from thermophysical modeling (Delbo' et al. 2015) on spin period (Fig. 7). As in the case of MBAs, there appears to be a significant trend to larger thermal inertia with increasing spin period.

Furthermore, a similar trend may also hold for Centaurs/TNOs, although the correlation with present data is not significant beyond the 1.6 $\sigma$ level (see Lellouch et al. 2013, fig. 8).

We suggest the explanation for the possibly universal trend of increasing thermal inertia with decreasing spin rate lies not in external influences, but rather in the different depths to which the thermal wave penetrates in otherwise similar asteroids rotating at different rates. The depth at which the amplitude of the diurnal thermal wave decays to 1/e of its surface value, known as the skin depth, is given by $d_s = (2\kappa/\rho c\omega)^{0.5} = (2/\omega)^{0.5} \Gamma/\rho c$ (Wesselink, 1948; Spencer et al. 1989). Evidence is accumulating that the uppermost surface layer of an asteroid is very different in terms of porosity, thermal conductivity and density compared to sub-surface layers just a few tens of centimeters below. Data collected at millimeter and submillimeter wavelengths from the MIRO radiometer/spectrometer on-board the Rosetta spacecraft, indicate that the main-belt asteroid 21 Lutetia has a highly insulating surface layer around 1 - 3 cm thick, with $\Gamma < 20$ J m$^{-2}$s$^{-0.5}$K$^{-1}$, on top of a "transition region" in which $\Gamma$ increases to 60 - 120 J m$^{-2}$s$^{-0.5}$K$^{-1}$ some 10 - 50 cm below the surface (Gulkis et al. 2012). It appears that the thermal conductivity and/or the material density, and therefore the thermal inertia, increase rapidly as depth increases. The lunar surface is thought to have a similar profile on the basis of data gathered during the Apollo missions, with a low-thermal-inertia porous surface layer 1 - 2 cm thick, covering a compacted sub-surface layer with some 50% higher density and an order of magnitude higher thermal conductivity at a depth of > 3 cm (Keihm and Langseth, 1975). These results imply that the rapid increase of density and thermal conductivity with depth can cause an increase of about a factor 4 or more in thermal inertia just a few centimeters below the surface.

## 4. IMPLICATIONS FOR ASTEROID REGOLITH STRUCTURE

The temperature distribution on the surface is largely determined by the thermal conductivity and density of material above the skin depth. Figure 8 is a plot of $\Gamma$ against skin depth for the MBA dataset used in Fig. 5 for bulk density values of $\rho$ = 1000 and 3000 kg m$^{-3}$, assuming $c$ = 680 J kg$^{-1}$ K$^{-1}$ (Chesley et al. 2003; Čapek, & Vokrouhlický, 2004); the density would be expected to increase with depth from the lower value, which is representative of porous surface material. Superimposed on the plot of Fig. 8 are the $\Gamma$ versus $d_s$ data for the NEOs of Fig. 7, for $c$ = 680 J kg$^{-1}$ K$^{-1}$ and $\rho$ = 3000 kg m$^{-3}$. It is evident that the thermal inertia rises rapidly with skin depth, consistent with the findings for the Moon and Lutetia (see above), and the rise is steeper for NEOs than for MBAs. Extrapolation of the trends in Fig. 8 suggest that thermal inertia values representative of solid rock (~ 2500 J m$^{-2}$s$^{-0.5}$K$^{-1}$, Golombek et al. 2003 and references therein, for Martian rocks with diameter > 20 cm) are reached some tens of centimeters to meters below the surface in the case of the MBAs (the median diameter in our dataset = 24 km). In the case of the much smaller (km-sized) NEOs our results indicate that the porous surface layer is thinner, and suggest that large pieces of solid rock exist just a meter or less below the surface. Note that the thermal inertia values derived from observational data are effective values relating mainly to the material layers above the skin depth, $ds$. Assuming thermal inertia increases with depth, the effective values should be considered lower limits for the thermal inertia at depth = $ds$.

## 5. SUMMARY AND CONCLUSIONS

We have developed a simple empirical relationship enabling the thermal inertia of an asteroid to be estimated given measurements of its thermal-infrared continuum and knowledge of its spin vector. The estimator provides thermal inertia values that agree with those from detailed thermophysical modeling for 34 objects with a RMS fractional deviation of 40%. The thermal inertia of asteroids, as determined by thermal-infrared observations, exhibits a strong dependence on spin rate, increasing as spin rate decreases. We interpret our discovery of spin-rate-dependent thermal inertia in terms of rapidly increasing material density and thermal conductivity with depth. Thermal inertia appears to increase by factors of 10 (MBAs) to 20 (NEOs) within a depth of just 10 cm. Our results are consistent with a very general picture of rapidly changing material properties in the topmost regolith layers of asteroids, as found in the case of 21 Lutetia (Gulkis et al. 2012). For a spin period > 10 h, knowledge of the rotation rate of an asteroid is crucial to choosing an appropriate value of thermal conductivity for calculation of the Yarkovsky effect; these results are also of relevance to modelers concerned with the mass and velocity distributions of ejecta expelled by a kinetic impactor spacecraft, and the corresponding ejecta-related momentum enhancement factor. Efforts should be made to carry out sophisticated thermophysical modeling of slowly-rotating asteroids, including depth-dependent thermal properties, for the purpose of probing the structure of the sub-surface material.

This publication makes use of data products from *WISE*/NEO*WISE*, which is a project of the Jet Propulsion Laboratory/California Institute of Technology, funded by the Planetary Science Division of the National Aeronautics and Space Administration. We gratefully acknowledge the excellent JPL Solar System Dynamics web service of which we have made extensive use, and the thoughtful comments of the anonymous referee. This work was supported by the European Union's Horizon 2020 research and innovation programme under grant agreement no. 640351 (project NEOShield-2).

FIGURES

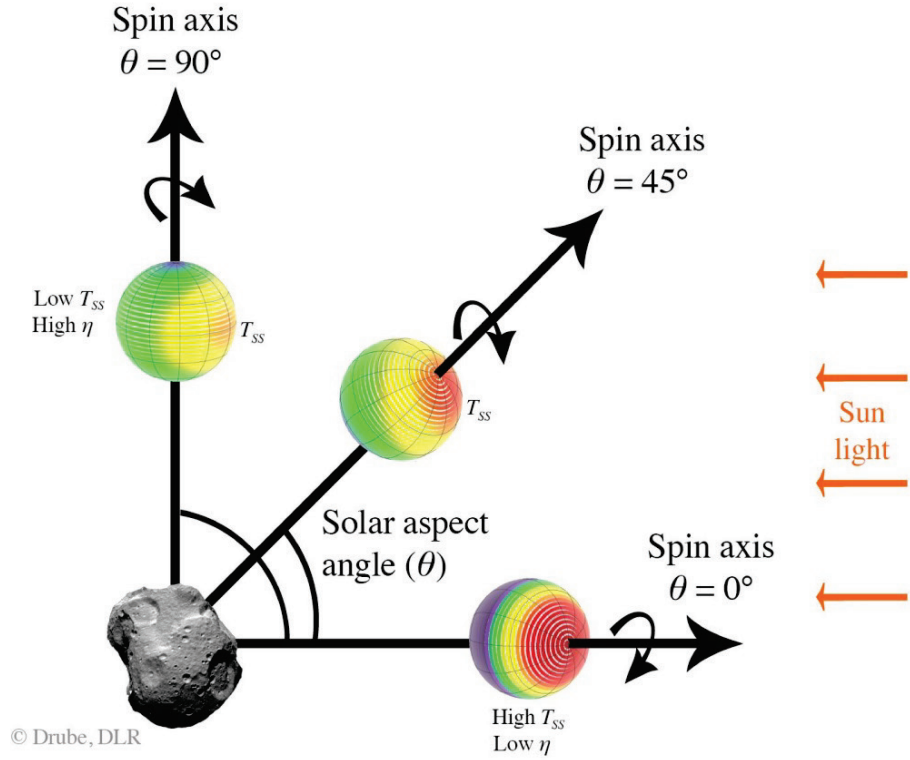

**Figure 1:** Illustration of the effect of the solar aspect angle, $\theta$, on the surface temperature distribution of a smooth-surface model asteroid. The solar aspect angle is the angle between the spin vector of the asteroid and the solar direction. The surface temperature distribution is governed by the rotation rate, the thermal inertia, and $\theta$. See text for definitions of the symbols.

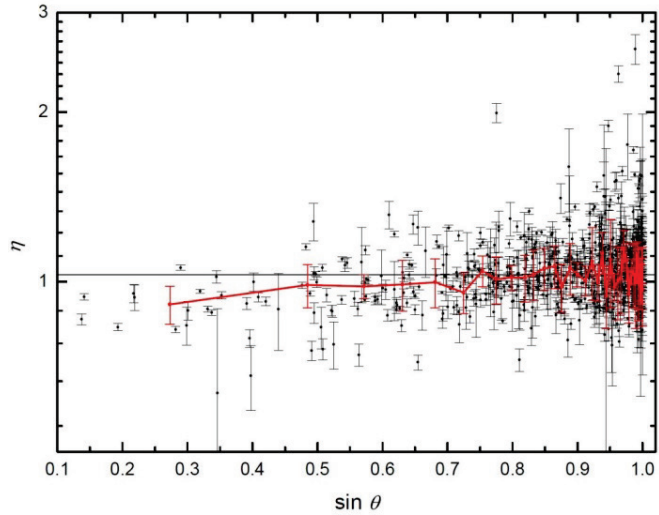

**Figure 2:** Confirmation that there is a general decrease in $\eta$ with decreasing solar aspect angle, $\theta$. Plotted values of $\eta$ (Masiero et al. 2011) are for MBAs with known spin vectors (outliers may be due to poor spin-axis determinations or $\eta$ values). The horizontal line represents the median of the plotted $\eta$ values. The red points and line trace the running weighted mean of the $\eta$ values (in bins of 20 points). Note that *no information on solar aspect angle* is used in generating the $\eta$ values published by the *WISE* project.

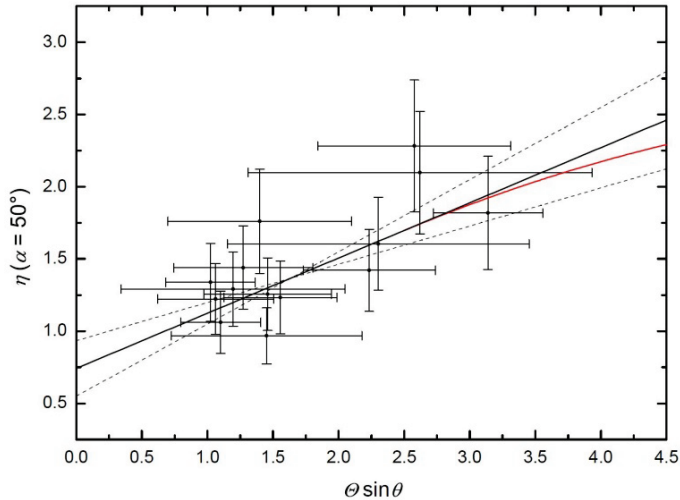

**Figure 3:** Normalized $\eta$ versus $\Theta \sin\theta$ for the NEOs in the compilation of Delbo' et al. (2015), where $\Theta$ is the thermal parameter and $\theta$ the solar aspect angle. The data set used here includes only those objects for which robust $\eta$ values could be obtained. The $\eta$ values have been normalized to a solar phase angle of 50°, as explained in the text. The continuous thick line is a weighted linear best fit given by $\eta_{norm} = 0.74 + 0.38 \times \Theta \sin\theta$; the dashed lines represent 1 $\sigma$ deviations from the best fit. The red curve is indicative of the form of the theoretical dependence of $\eta$ on $\Theta$ (it is not a formal fit). Independent measurements of $\eta$ for the same object are included as separate data points. The dataset used is given in Table 1 (the fractional uncertainties in $\Theta \sin\theta$ derive from those in $\Gamma_{TP}$).

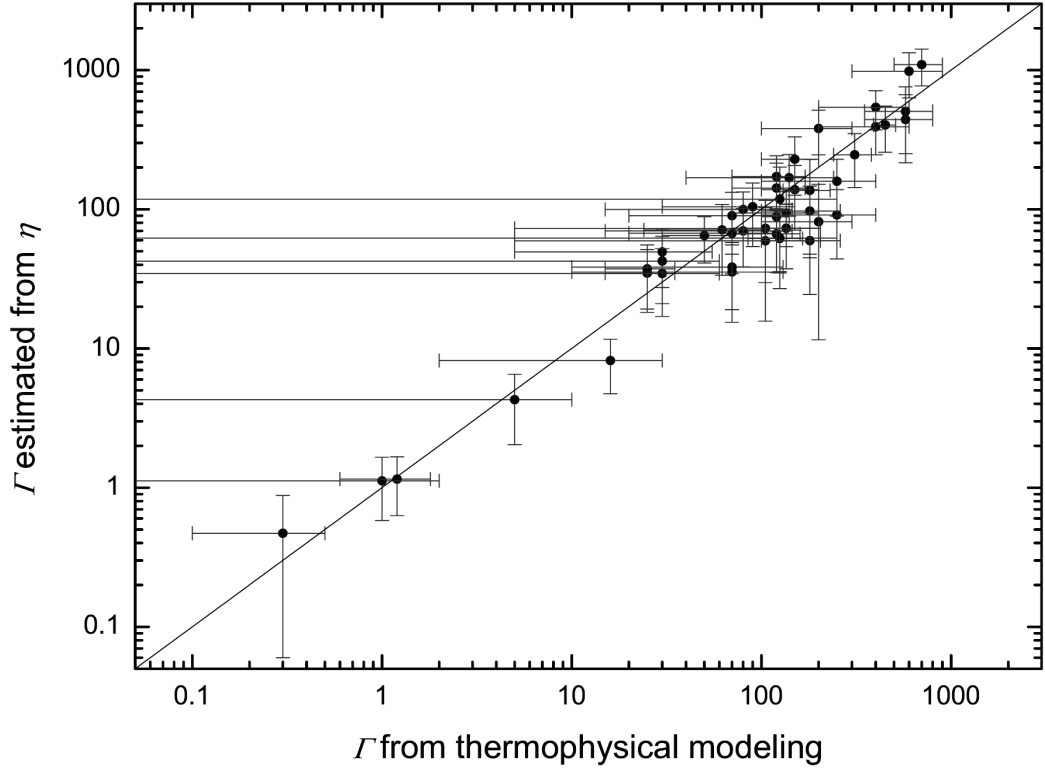

**Figure 4:** Test of the NEATM-based thermal-inertia estimator (Equation 2). Estimated values of thermal inertia, $\Gamma$ [J m$^{-2}$s$^{-0.5}$K$^{-1}$], are plotted against $\Gamma$ derived by means of detailed thermophysical modeling. The data set excludes objects with $\Theta \sin\theta$ outside the range 0.75 – 3.5 (see text). The error bars on the y-axis result from the assumption of uncertainties in $\eta$ of ± 20%. There is good agreement between the two sets of values over nearly 4 orders of magnitude in thermal inertia. As in Fig. 3 values of $\eta$ for an object derived from independent sets of data are treated as separate values, thus some objects are represented by two data points. The RMS fractional deviation, $(\Gamma_{TP} - \Gamma_{est})/\Gamma_{TP}$, is 40%, where $\Gamma_{TP}$ and $\Gamma_{est}$ refer to thermal inertia derived from thermophysical modeling and thermal inertia estimated from $\eta$, respectively. See Table 1 for the data plotted and associated parameter values.

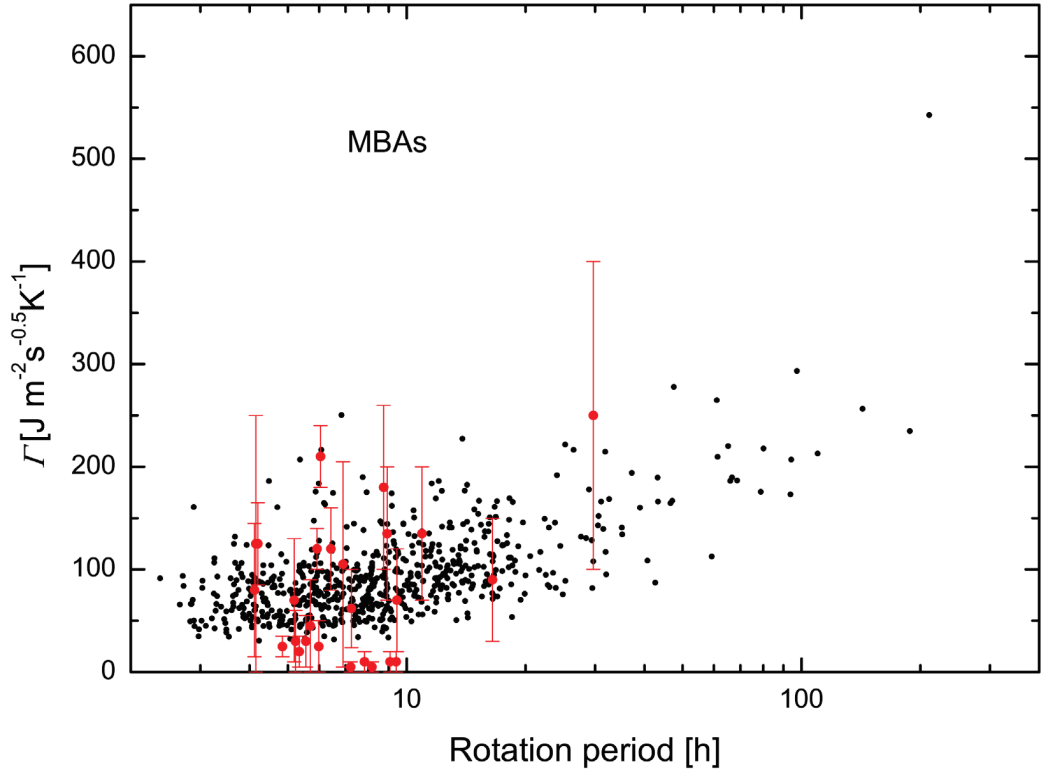

**Figure 5:** Estimated thermal inertia versus rotation period for MBAs. The NEATM-based thermal-inertia estimator (Equation 2) was used to estimate values of $\Gamma$ from $\eta$ values given in the *WISE* catalog of Masiero et al. (2011) for objects with known spin vectors (black points; note that the data set excludes objects with $\Theta \sin\theta$ outside the range 0.75 – 3.5, such as those with very low thermal inertia). There is a clear trend to higher values of thermal inertia for rotation period > 10 h. Error bars have been omitted for clarity. Uncertainties of ± 20% in the *WISE* $\eta$ values result in a mean fractional uncertainty of ± 47% for the plotted thermal inertia values. The median diameter of the MBAs in the dataset is 24 km. Available thermal inertia values from detailed thermophysical modeling (Delbo' et al. 2015) are superimposed for comparison (red points with error bars).

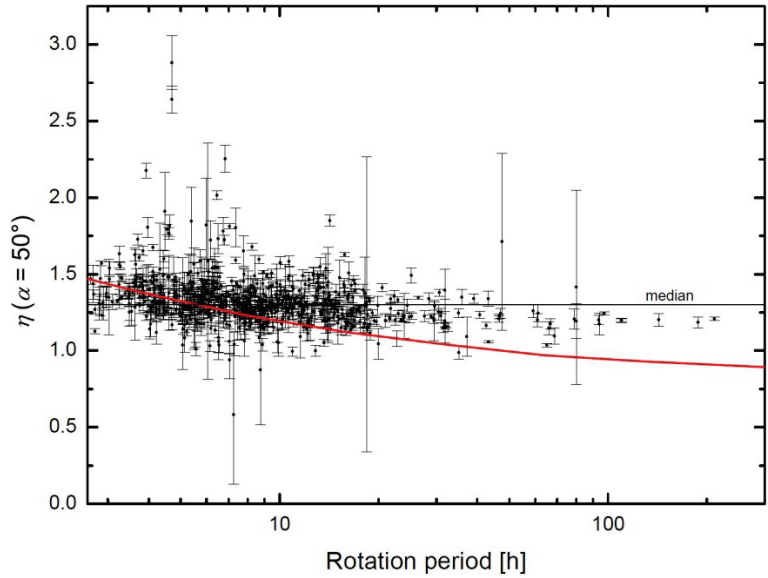

**Figure 6:** $\eta$ versus rotation period. The red continuous curve represents the expected relation between $\eta$ and rotation period on the basis of a smooth-surface thermophysical model based on spherical geometry for a constant thermal inertia of 75 J m$^{-2}$s$^{-0.5}$K$^{-1}$ (cf. Fig. 5). The $\eta$ values have been normalized to a solar phase angle of 50°, as explained in the text. The curve is normalized at a rotation period of 4 h to $\eta$ =1.37, the median value in the range 3.0 – 5.0 h. The horizontal line represents the median of the plotted $\eta$ values. As rotation period increases the $\eta$ values remain relatively high, consistent with increasing thermal inertia.

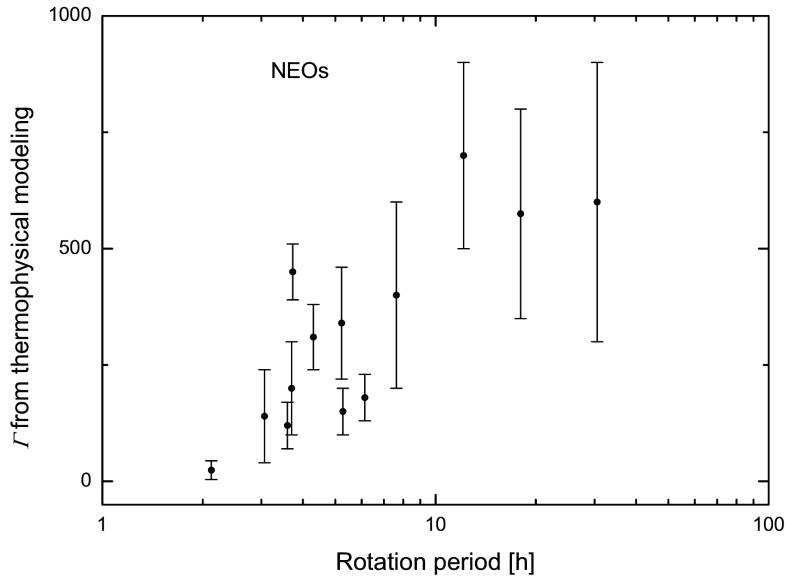

**Figure 7:** Thermal inertia (J m$^{-2}$s$^{-0.5}$K$^{-1}$) versus rotation period for NEOs in the dataset of Delbo' et al. (2015). As in the case of MBAs, slowly-rotating NEOs appear to be associated with higher values of thermal inertia (note: 54509 YORP, which has a very short rotation period of 0.20 h, is not shown in this plot).

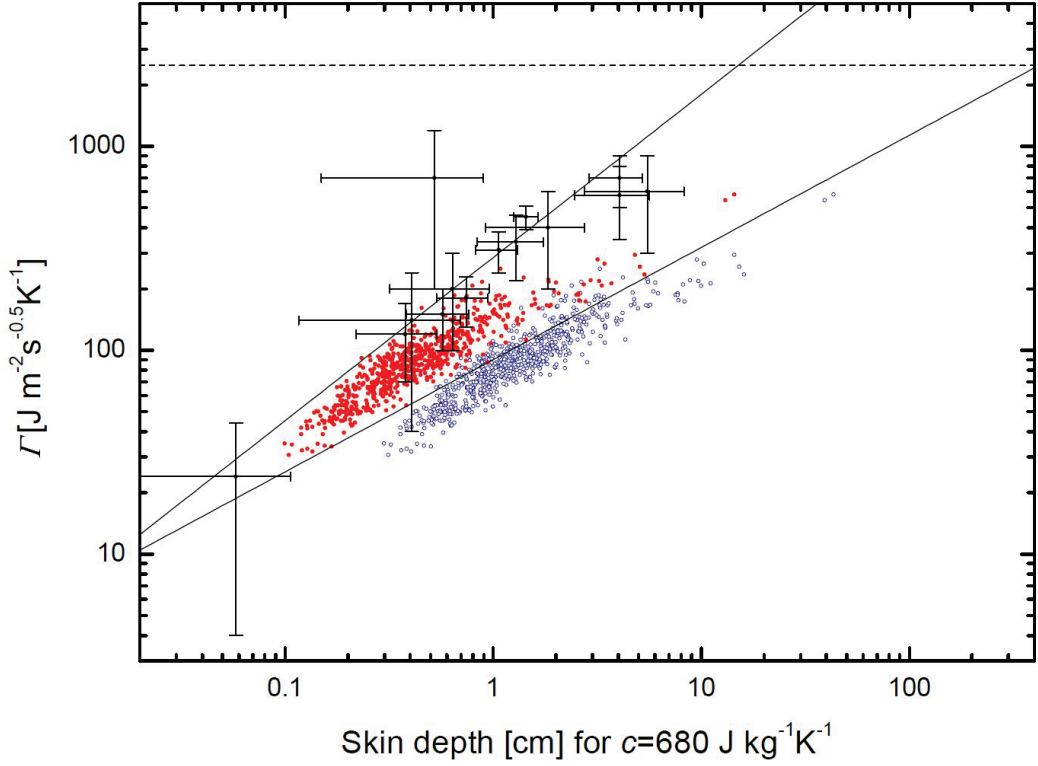

**Figure 8:** Thermal inertia versus skin depth. The datasets are those used for Figs. 5 and 7. In the case of the MBAs $\Gamma$, from the NEATM-based thermal-inertia estimator, is plotted for bulk density values of 1000 kg m$^{-3}$ (open blue circles) and 3000 kg m$^{-3}$ (filled red circles), assuming $c$ = 680 J kg$^{-1}$K$^{-1}$; error bars have been omitted for clarity (see caption to Fig. 5). The continuous lines represent the envelope of the data set for $\rho$ = 3000 kg m$^{-3}$. The dashed horizontal line at $\Gamma$ = 2500 J m$^{-2}$s$^{-0.5}$K$^{-1}$ represents the thermal inertia of solid rock. The NEO data of Fig. 7 (from thermophysical modeling) are superimposed (black points with error bars), taking $c$ = 680 J kg$^{-1}$ K$^{-1}$ and $\rho$ = 3000 kg m$^{-3}$. Note that values of thermal inertia can be converted to thermal conductivity by substitution of the assumed values of $\rho$ and c in the expression $\kappa = \Gamma^2/\rho c$; for reference, taking $\rho$ = 2000 kg m$^{-3}$, $c$ = 680 J kg$^{-1}$K$^{-1}$, the values of thermal conductivity corresponding to dusty lunar-like regolith (~50 J m$^{-2}$s$^{-0.5}$K$^{-1}$) and solid rock (~2500 J m$^{-2}$s$^{-0.5}$K$^{-1}$) are 0.002 Wm$^{-1}$K$^{-1}$ and 5 Wm$^{-1}$K$^{-1}$, respectively.